\documentclass[preprint,showpacs,nofootinbib,preprintnumbers,amsmath,amssymb]{revtex4}

\usepackage{graphicx}
\usepackage{dcolumn}
\usepackage{bm}

\begin{document}


\title{Podolsky Electromagnetism at Finite Temperature: Implications on Stefan-Boltzmann Law}

\author{C. A. Bonin}
 \email{bonin@ift.unesp.br}
\author{R. Bufalo}%
 \email{rbufalo@ift.unesp.br}
 \author{B. M. Pimentel}%
 \email{pimentel@ift.unesp.br}
 \affiliation{%
Instituto de F\'{i}sica Te\'{o}rica UNESP - S\~{a}o Paulo State University.\\
Caixa Postal 70532-2, 01156-970 S\~{a}o Paulo, SP, Brazil.
}%
\author{G. E. R. Zambrano }%
 \email{gramos@udenar.edu.co}
  \affiliation{%
Departamento de F\'{i}sica, Universidad de Nari\~{n}o\\
Clle 18 Cra 50, San Juan de Pasto, Nari\~{n}o, Colombia
}%

\date{\today}

\begin{abstract}
In this work we study Podolsky electromagnetism  in thermodynamic equilibrium. We show that a Podolsky mass-dependent modification to the Stefan-Boltzmann law is induced and we use experimental data to limit the possible values for this free parameter.
\end{abstract}

\maketitle

\section{Introduction}

Electromagnetism is a $U(1)$ gauge theory \cite{Quigg}. The electromagnetic field emerges from the application of the gauge principle to the local Abelian group. This principle tells us that the electromagnetic field must be a Lorentz vector with some well-defined internal properties. What the principle does not tell us is which is the highest order of the field derivatives appearing in the Lagrangian of the theory. Usually, we just invoke \textit{Occam's razor} and use the simplest one: a Lagrangian with only first-order derivatives. It is well known that there exist Lagrangian densities of second-order derivatives which are equivalent to Lagrangian densities with derivatives of only first order. For instance, consider the following two Lagrangian densities for the real scalar field: $\mathcal{L}_1=\frac{1}{2}\partial^\mu\phi\partial_\mu\phi+V(\phi)$ and $\mathcal{L}_2=-\frac{1}{2}\phi\square\phi+V(\phi)$. They both lead to the same equation of motion \cite{Higher Order Scalar}. Since they lead to the very same physical consequences, we are free to choose either one or the other. This situation is unlike what occurs with the electromagnetic field. It has been proved that there is only one extension (up to a total-divergence term) for a second-order derivative Lagrangian for electromagnetism that is both Lorentz and gauge-invariant \cite{Cuzinatto_1}. This extension is known as \textit{Podolsky electromagnetism}, after Boris Podolsky has first proposed it in 1942 \cite{Podolsky_1}. As we shall review in the next section, Podolsky Lagrangian does not lead to the equations of motion expected from the Maxwell theory. Therefore, they are \textit{non-equivalent}  descriptions of the Abelian gauge field.


Podolsky electromagnetism depends on a free parameter. The way used to fix the Podolsky parameter is the same one used to set bounds on the values of all free parameters of the Standard Model of Elementary Particles: it can only be fixed from experiments \cite{Cuzzinatto_2}. Indeed, this is one of the aims of this paper. Besides, despite its long-dated success, Maxwell electromagnetism still has unsolved problems. In the classical level, for example, the electrostatic potential diverges over punctual electric charges. On the other hand, Podolsky's electrostatic potential is finite everywhere \cite{Podolsky_1}. In addition, using Podolsky theory, Frenkel was able to solve the famous ``4/3 problem of classical electrodynamics" \cite{Frenkel}, while in the context of Maxwell theory this problem remains open.  Furthermore, Podolsky theory of electromagnetic interaction presents a richer theoretical structure than a theory with only first-order derivatives. Finally, with Podolsky electromagnetism there is the possibility of new Physics. We see, then, that Podolsky theory is interesting from both theoretical and experimental points of view.

In 1901, Planck's pioneer and now famous work on black body radiation led to the foundations of Quantum Physics. In order to fit the available data, Planck postulated that the energy exchanging between cavity oscillators and the Maxwell electromagnetic field in thermal equilibrium is quantized. Under these assumptions he derived the frequency distribution of the black body radiation. In his work, Planck was also able to deduce the Stefan-Boltzmann law, which states that the power for unit area of the black body radiation grows with the fourth power of the temperature. We can cite as an example of Planckian black body distribution the Cosmic Microwave Background Radiation (CMBR) \cite{Penzias and Wilson}. In fact, with precise data, CMBR was considered the most accurate black body radiation  measured up to date \cite{White}.  Nowadays, Planck's law can be derived using modern methods of Quantum Field Theory at Finite Temperature \cite{Le Bellac, Das, Kapusta}. In this context, black body radiation is seen as a gas of Maxwell photons in thermodynamic equilibrium. In the light of this thought, in the present work we investigate a gas of Podolsky photons at thermal equilibrium and we seek the finite-temperature properties of the Podolsky theory.  Through use of imaginary-time technique from Finite-Temperature Field Theory we construct the partition function of the theory after determining its constraint structure. From the partition function we are able to evaluate all thermodynamical quantities, including the energy density distribution. We expect a modification on the Stefan-Boltzmann law  due to the presence of the term that contains higher-order derivatives  in the Podolsky Lagrangian density. Finaly, we compare our results with experimental data for the Stefan-Boltzmann law at CMBR temperature and we set a thermodynamical limit on the Podolsky parameter.

This paper is organized as follows. In section \ref{canonical} we review the canonical structure of Podolsky electromagnetism. In section \ref{path integral} we work on the path integral formulation of the transition amplitude which is a step to the attainment of the Partition Function of the problem.  In section \ref{finite temperature} we use imaginary-time formalism and deal with the problem of evaluating the Partition Function. In that section we also show the Podolsky correction to the Stefan-Boltzmann law and we use experimental data to limit the free parameter of the theory. Our final remarks are presented in section \ref{conclusions}.

\section{General aspects of  Podolsky Theory's canonical Structure}\label{canonical}

The content of this section is by no means new. We intend only to review some general aspects of the canonical structure of Podolsky electromagnetism.

The Lagrangian density for Podolsky theory is\footnote{Throughout this work we set natural unit system, the metric signature $(+---)$, and $\square\equiv\partial^\mu\partial_\mu$ as long as we work in the Minskowski spacetime.} \cite{Podolsky_1}
\begin{align}
\mathcal{L}=-\frac{1}{4}F_{\mu\nu}F^{\mu\nu}+\frac{1}{2m^2}\partial_\mu F^{\mu\nu}\partial_\xi F^{\xi}_{\phantom{\xi}\nu},\label{lagrangian}
\end{align}
where, like in Maxwell's case, the field-strength tensor is $F_{\mu\nu}=\partial_\mu A_\nu-\partial_\nu A_\mu$. The free parameter $m$ has dimension of energy and in the limit $|m|\rightarrow\infty$ we recover the Maxwell Lagrangian density. The Euler-Lagrange equations for this theory are

\begin{align}
\left(\square+m^2\right)\partial_\mu F^{\mu\nu}=0.\label{eq. of motion}
\end{align}

As we have anticipated in the introductory section, these equations of motion differ from those of Maxwell theory. Therefore, the Physical contents of the Podolsky theory do not coincide with those of Maxwell theory.

The symmetric Energy-Momentum density tensor reads \cite{Barut}:

\begin{align}
\mathcal{T}^{\mu\nu}=&\,F^\mu_{\phantom{\mu}\lambda}F^{\lambda\nu}-\eta^{\mu\nu}\mathcal{L}+\frac{1}{m^2}\left(2\partial^\lambda F^{\mu\xi}\partial_\lambda F^\nu_{\phantom{\nu}\xi}-2\partial^\lambda F^{\xi\mu}\partial_\xi F_\lambda^{\phantom{\lambda}\nu}+\partial_\lambda F^{\lambda\mu}\partial_\xi F^{\xi\nu}\right).
\end{align}

The energy density $\mathcal{E}$ is the component $00$ of this tensor. It is possible to write $\mathcal{E}$ in terms of the electric and magnetic fields:

\begin{align}
\mathcal{E}=\frac{1}{2}\left\{\mathbf{E}^2+\mathbf{B}^2+\frac{1}{m^2}\left[\left(\nabla\cdot\mathbf{E}\right)^2 +\left(\dot{\mathbf{E}}-\nabla\times\mathbf{B}\right)^2+4\mathbf{E}\cdot\square\mathbf{E}+4\mathbf{E}\cdot\nabla\left(\nabla\cdot\mathbf{E}\right)\right]\right\}.
\end{align}

This expression does not appear to be positive-definite in the general case. However, if we restrict it to the electrostatic case, we have for the energy \cite{Podolsky_1}:

\begin{align}
E_{electrost.}=&\,\int d^3 x\, \mathcal{E}_{electrost.}=\frac{1}{2}\int \hspace{-.1cm} d^3x\hspace{-.1cm}\left[\mathbf{E}^2+\frac{1}{m^2}\left(\nabla\cdot\mathbf{E}\right)^2\right]. \label{positive energy}
\end{align}

Once we impose the condition $E_{electrost.}\geq 0$, we have the implication that the parameter $m$ must be real and, without loss of generality, we assume it to be positive.

Now, we can impose the \textit{generalized Lorenz condition} $\left(\square+m^2\right)\partial_\mu A^\mu=0$ on the Podolsky field and the equations of motion (\ref{eq. of motion}) become simplified \cite{Pimentel}:

\begin{align}
\left(\square+m^2\right)\square A^\mu=0.
\end{align}

One possible solution to this equation can be written as

\begin{align}
A^\mu =A^\mu_{M}+A^\mu_{P},\label{solution}
\end{align}
where $A^\mu_{M}$ satisfy the equations of motion of Maxwell electromagnetism $\square A^\mu_{M}=0$ and $A^\mu_{P}$ satisfies the Proca equations $\left(\square+m^2\right) A^\mu_{P}=0$. Here we notice a notable difference between Maxwell and Podolsky theories. Although Maxwell field has only four components (one for each spacetime direction), Podolsky's has eight. For each direction the Podolsky vector field has one massless and one massive \textit{sectors}, as it is seen in (\ref{solution}). This interpretation will be useful in the next sections.

We can write the canonical Hamiltonian of Podolsky theory $H_C$ as

\begin{align}
H_C=\int d^3x\left(p_\mu\dot{A}^\mu+\pi_\mu\ddot{A}^\mu-\mathcal{L}\right),
\end{align}
with the momenta defined as

\begin{align}
p_\mu\equiv&\, \frac{\partial\mathcal{L}}{\partial\left(\dot{A}^\mu\right)} -\partial_0\left[\frac{\partial\mathcal{L}}{\partial\left(\ddot{A}^\mu\right)}\right] -\partial_k\left[\frac{\partial\mathcal{L}}{\partial\left(\partial_0\partial_k A^\mu\right)} +\frac{\partial\mathcal{L}}{\partial\left(\partial_k \dot{A}^\mu\right)}\right];\\
\pi_\mu\equiv &\,\frac{\partial\mathcal{L}}{\partial\left(\ddot{A}^\mu\right)},
\end{align}
which are canonically conjugated respectively to the fields $A^\mu$
and $\varphi^\mu\equiv \dot{A}^\mu$, which is considered  as
an independent variable.

Expliciting these quantities $\left(H_C=\int d^3x\mathcal{H}_C\right)$:

\begin{align}
p_\mu=&\,-F_{0\mu}+\frac{1}{m^2}\left(\eta^k_{\phantom{k}\mu}\partial_k\partial_jF^{j0}-\partial_0\partial_\nu F^{\nu}_{\phantom{\nu}\mu}\right);\\
\pi_\mu=&\,\frac{1}{m^2}\left(\partial_\nu F^{\nu}_{\phantom{\nu}\mu}-\eta^0_{\phantom{0}\mu}\partial_k F^k_{\phantom{k}\mu}\right);\\
\mathcal{H}_C=&\,-\mathbf{p}\cdot\overrightarrow{\varphi}-\frac{1}{2}m^2\overrightarrow{\pi}^2+\pi^k\partial_j F^{jk}-\frac{1}{2}\left(\overrightarrow{\varphi}+\nabla A^0\right)^2+\nonumber\\
&\,-\frac{1}{2m^2}\left(\nabla^2A^0+\nabla\cdot\overrightarrow{\varphi}\right)^2+ \frac{1}{4}F^{jk}F^{jk}.
\end{align}

Now, we can follow the steps of reference \cite{Pimentel} for the constraint analysis \textit{a la} Dirac \cite{Dirac} and find that there are three (unlike only two found in Maxwell theory) first class constraints:

\begin{align}
\phi_1\equiv&\, \pi_0\approx 0;\\
\phi_2\equiv&\, p_0 -\nabla\cdot \overrightarrow{\pi}\approx 0;\\
\phi_3\equiv&\,\nabla\cdot\mathbf{p}\approx 0,
\end{align}
with the usual notation the symbol ``$\approx$" means \textit{weak equality}. Following Dirac's procedure, we choose three gauge conditions:

\begin{align}
\Omega_1\equiv&\, \varphi^0\approx 0;\label{omega 1}\\
\Omega_2\equiv&\,\left(\square+m^2\right)\nabla\cdot\mathbf{A}\approx 0;\\
\Omega_3\equiv&\, A^0\approx 0.\label{omega 3}
\end{align}

It was shown in \cite{Pimentel} that this set constitutes an
appropriated  non-covariant gauge condition which fixes the
first-class constraints. Solution (\ref{solution}) has made it clear
that Podolsky electromagnetic field has eight apparently independent
components. However, gauge conditions (\ref{omega 1}-\ref{omega 3})
reduces this number to five. Five ``degrees of freedom" (d. o. f.)
are compatible with the interpretation that Podolsky field is
composed of a Maxwell field (with two d. o. f.) plus a Proca field
(which has three d. o. f.).

\section{Path Integral Formalism}\label{path integral}

Once constraint analysis is done, we are able to write down the generating functional with null sources (or transition amplitude):

\begin{align}
\tilde{Z}=&\,\int Dp_\sigma D\pi_\varsigma DA^\sigma D\varphi^\varsigma \det \left\{\Omega_a,\phi_b\right\}\left[\prod_{n=1}^3\delta\left[\phi_n\right]\delta\left[\Omega_n\right]\right] \exp\left(i\int d^4x \mathcal{L}_C\right)\label{why not}
\end{align}
where

\begin{align}
\det \left\{\Omega_a,\phi_b\right\}=&\,\det \left[\left(\square+m^2\right)\nabla^2\right];\\
\mathcal{L}_C=&\, p_\mu\partial_tA^\mu+\pi_\mu\partial_t\varphi^\mu-\mathcal{H}_C.
\end{align}

After some steps not quite different from those of Maxwell's case, we can rewrite the transition amplitude as

\begin{align}
\tilde{Z}=&\,\int \left[\prod_{\sigma=0}^3DA^\sigma\right]\det\left[\left(\square+m^2\right)\nabla^2\right] \delta\left[\left(\square+m^2\right)\nabla\cdot\mathbf{A}\right]\exp\left(i\int d^4x\mathcal{L}\right)
\end{align}
with $\mathcal{L}$ given by equation (\ref{lagrangian}).

Using a straight-forward generalization of the Faddeev-Popov \textit{ans\"{a}tz}, we can pass from a non-covariant gauge fixing to a covariant one:

\begin{align}
\tilde{Z}=&\,\int \left[\prod_{\sigma=0}^3DA^\sigma\right]\det\left[\frac{1}{\rho}\left(\square+m^2\right)\square\right] \delta\left[\frac{1}{\rho}\left(\square+m^2\right)\partial_\varsigma A^\varsigma - f\right]\exp\left(i\int d^4x\mathcal{L}\right),
\end{align}
where $\rho\neq 0$ is an arbitrary real number and $f=f(x)$ is an arbitrary real function. Physical quantities are independent of the function $f(x)$. So, we ``sum" $\tilde{Z}$ over all functions $f$, considering the weight factor $\exp\left(-\frac{i}{2}\int d^4x f^2\right)$:

\begin{align}
\bar{Z}\equiv&\, \int Df \tilde{Z}\exp\left(-\frac{i}{2}\int d^4x f^2\right)\nonumber\\
=&\, \int D\bar{c} Dc  \left[\prod_{\sigma=0}^3DA^\sigma\right]\exp\left(i\int d^4x \mathcal{L}_{eff}\right),\label{transition}
\end{align}
where $\bar{c}$ and $c$ are  ghost fields. The \textit{effective} Lagrangian density $\mathcal{L}_{eff}$ is defined as\

\begin{widetext}
\begin{align}
\mathcal{L}_{eff}\equiv -\frac{1}{4}F_{\mu\nu}F^{\mu\nu}+\frac{1}{2m^2}\partial_\mu F^{\mu\nu}\partial_\xi F^{\xi}_{\phantom{\xi}\nu}-\frac{1}{2\rho^2}\left[\left(\square+m^2\right)\partial_\mu A^\mu\right]^2-\frac{1}{\rho}\bar{c}\left(\square+m^2\right)\square c.
\end{align}
\end{widetext}

So far we have studied Podolsky theory at zero temperature. In the next section we will analyze the theory in thermodynamic equilibrium.

\section{Finite Temperature}\label{finite temperature}

It is possible to obtain the Partition Function from the transition amplitude (\ref{transition}). Once the Partition Function is carried out, all thermodynamical properties of the system becomes available. In order to obtain the Partition Function from the transition amplitude we just have to perform an Euclideanization of the time components of the vector fields, a compactification of the Wick-rotated time coordinate and impose periodic boundary conditions ($P$) in this coordinate for the electromagnetic and the ghost fields \cite{Das}.\footnote{The Euclideanization follows as $x^0\rightarrow -i\tau$ (and therefore $\partial_0\rightarrow i\partial/\partial\tau\equiv i\partial_\tau$) and $A^0\rightarrow -iA_0^E\equiv-iA_0$. All summations are performed with Euclidean metric from now on \cite{De Witt}.} Doing so, we find the Partition Function for the free Podolsky field:

\begin{align}
Z(\beta)=&\, \int_{P} D\bar{c} Dc  \left[\prod_{\sigma=0}^3DA_\sigma\right]\exp\left(-\int_\beta dx \mathcal{L}_{E}\right),
\end{align}
where $\beta\equiv 1/T$, $T$ is the temperature, and we use the notation

\begin{align}
\int_{\beta} dx\equiv&\, \int_0^\beta d\tau\int d^3x;\\
\Delta\equiv&\, -\partial_\gamma\partial_\gamma;\\
\mathcal{L}_E\equiv&\, -\frac{1}{4}F_{\zeta\lambda}F_{\zeta\lambda}-\frac{1}{2m^2}\partial_\zeta F_{\zeta\lambda}\partial_\xi F_{\xi\lambda}+\nonumber\\
&\,-\frac{1}{2\rho^2}\left[\left(\Delta+m^2\right)\partial_\lambda A_\lambda\right]^2-\frac{1}{\rho}\bar{c}\left(\Delta+m^2\right)\Delta c.
\end{align}

$\mathcal{L}_E$ is called the effective Euclidean Lagrangian density. Since there is no coupling among the electromagnetic field and the ghost fields, the Partition Function takes the form

\begin{align}
Z\left(\beta\right)=&\,\int_{P}D\bar{c}Dc \exp\left\{-\int_\beta dx\left[\frac{1}{\rho}\bar{c}\left(\Delta+m^2\right)\Delta c\right]\right\}\times\nonumber\\
&\,\times\int_P \left[\prod_{\sigma=0}^3DA_\sigma\right]\exp\left(-\frac{1}{2}\int_\beta dx A_\alpha M_{\alpha\gamma} A_\gamma\right)\nonumber\\
=&\,\det\left[\frac{1}{\rho}\left(\Delta+m^2\right)\Delta\right]\left[\mbox{Det}\left(M_{\alpha\gamma}\right)\right]^{-\frac{1}{2}},
\end{align}
where ``$\det$" stands for the determinant in the Hilbert space alone (as usual) and ``Det" stands for the determinant in both Euclidean spacetime and the Hilbert space. The operator $M_{\alpha\gamma}$ is defined as

\begin{align}
M_{\alpha\gamma}\equiv&\,\left(1+\frac{\Delta}{m^2}\right)\Delta\delta_{\alpha\gamma}+\left(1+\frac{\Delta}{m^2}\right)\left[1 +\left(\frac{m^2}{\rho}\right)^2\left(1+\frac{\Delta}{m^2}\right)\right]\partial_\alpha\partial_\gamma.
\end{align}

After evaluating the determinant in the Euclidean spacetime we find

\begin{align}
Z\left(\beta\right)=\left[\det\left(\Delta\right)\right]^{-1}\left[\det\left(\Delta+m^2\right)\right]^{-\frac{3}{2}}.
\end{align}

We note that the Partition Function is a product of determinants of the form $\left[\det\left(\Delta+m_j^2\right)\right]^{-\frac{n_j}{2}}$, with $j=1$, and $2$. Each of these terms describes a gas of free particles with mass $m_j$ and $n_j$ d. o. f. We identify the first of these determinants as a partition function for massless particles with two d. o. f., \textit{i. e.}, Maxwell photons. On the other hand, the second determinant is the partition function for particles of mass $m$ and three d. o. f. Those are Proca particles. Since $Z\left(\beta\right)$  involves no other terms, it describes a gas formed of \textit{non-interacting} gases of free Maxwell photons and free Proca bosons. The ``non-interaction" property between the two distinct gases is a direct consequence of Podolsky theory's linearity.

In order to evaluate the determinants we note that the equation

\begin{align}
\det\left(\Delta+m^2\right)=\prod_{n,\mathbf{p}}\beta^2\left[\omega_n^2+\omega^2\left(p,m\right)\right],
\end{align}
with the relativistic energy-momentum relation $\omega\left(p,m\right)\equiv\sqrt{\mathbf{p}^2+m^2}$ and the bosonic Matsubara frequencies $\omega_n\equiv 2n\pi/\beta$,
 remains valid for both massive and massless cases \cite{Kapusta}. Using this identity, the logarithm of the Partition Function reads

\begin{align}
\ln\left[Z\left(\beta\right)\right]=&-\sum_{n,\mathbf{p}}\ln\left[\beta^2\left(\omega_n^2+\mathbf{p}^2\right)\right] -\frac{3}{2}\sum_{n,\mathbf{p}}\ln\left\{\beta^2\left[\omega_n^2+\omega^2\left(p,m\right)\right]\right\}.
\end{align}

After evaluating the sum in $n$, passing to continuous in momentum space, and discharging irrelevant $\beta$-independent terms and vacuum contributions, we have:

\begin{widetext}
\begin{align}
\ln\left[Z\left(\beta,V\right)\right]=-2V\int\frac{d^3 p}{\left(2\pi\right)^3} \ln\left(1-e^{-\beta p}\right) -3V\int\frac{d^3 p}{\left(2\pi\right)^3} \ln\left[1-e^{\beta\omega\left(p,m\right)}\right].\label{z}
\end{align}
\end{widetext}

The first term in the r. h. s. of (\ref{z}) is associated with a gas of free Maxwell  photons and it gives the Planck's law. Since it can be found in many text books we shall skip its computation. The second term, on the other hand, corresponds to the massive sector of the theory. This term depends on the Podolsky parameter and shall give a correction to the Stefan-Boltzmann law. Calling the $m$-dependent term as $\ln\left(Z'\right)$ we find, after changing the integration variable to $x\equiv \omega/m$:\footnote{The sum over $k$ arises from the expansion of $\left(1-e^{-\beta\omega}\right)^{-1}.$}

\begin{align}
\ln\left(Z'\right)=\frac{m^4\beta V}{2\pi^2} \sum_{k=1}^\infty\int_1^\infty dx\left(x^2-1\right)^{2-\frac{1}{2}}e^{-k\beta mx}.
\end{align}

In order to write this expression in a compact form we use the following representation of the Modified Bessel Function of the Second Kind, valid for $n>-1/2$ \cite{Arfken}:

\begin{align}
K_n(z)=\frac{\sqrt{\pi}}{\Gamma\left(n+\frac{1}{2}\right)}\left(\frac{z}{2}\right)^n\int_1^\infty e^{-zx}\left(x^2-1\right)^{n-\frac{1}{2}} dx,
\end{align}
where $\Gamma(y)$ is the Gamma Function. We can now write

\begin{align}
\ln\left(Z'\right)=\frac{3}{2}\frac{\beta m^4 V}{\pi^2}\sum_{k=1}^\infty\frac{K_2\left(k\beta m\right)}{\left(k\beta m\right)^2}.\label{sum}
\end{align}

As far as we know, there is no known analytical, closed form to the summation appearing in this equation. Hence, we restrict ourselves to evaluate $\ln\left(Z'\right)$ only approximately.  We recall  that in the limit $m\rightarrow \infty$ the results found using Podolsky electromagnetism must go to the usual results of Maxwell theory. For this reason, we expect $\ln Z'$ to be a correction to the Planck law. In thermodynamic equilibrium, temperature is a natural scale of energy. In this sense, we will evaluate the correction to Stefan-Boltzmann law in the regime $\beta m=m/T\gg 1$. So, we can write

\begin{align}
K_2\left(k\beta m\right)\sim \sqrt{\frac{\pi}{2k\beta m}}e^{-k\beta m}
\end{align}
and keep only the first term of the sum in (\ref{sum}). Within this approximation, we can solve equation (\ref{z}):

\begin{align}
\ln\left[Z\left(\beta,V;m\right)\right]=\frac{\pi^2}{45}\frac{V}{\beta^{3}}+3V\left(\frac{m}{2\pi\beta}\right)^{\frac{3}{2}}e^{-\beta m}.\label{ln de z}
\end{align}

The first term in the r. h. s. of this equation is the usual Planck result. The second term is a correction due to the Podolsky mass. This equation enables us to evaluate any thermodynamical quantity. The energy density $u\left(T;m\right)$, for instance, is found to be

\begin{align}
u\left(T;m\right)=-\left.\frac{1}{V}\frac{\partial\ln\left[Z\left(\beta,V;m\right)\right]}{\partial \beta}\right|_{V}
=\sigma\left(T,m\right)T^4,\label{modified sb law}
\end{align}
with $\sigma\left(T,m\right)=\sigma_0+\delta\sigma\left(m/T\right)$, where $\sigma_0=\pi^2/15$ is the Stefan-Boltzmann constant and
\begin{align}
\delta\sigma\left(\frac{m}{T}\right)=&\frac{45\sigma_0}{\sqrt{8\pi^7}}\left(\frac{m}{T}\right)^{\frac{5}{2}}e^{-\frac{m}{T}}\label{sb}
\end{align}
is the correction due to the Podolsky parameter. As we can see, in the limit $m/T\rightarrow\infty$, $\sigma\left(T,m\right)$ goes to $\sigma_0$ and we recover the Stefan-Boltzmann law in (\ref{modified sb law}) as expected.

Equation (\ref{sb}) can be used to set a limit on the possible values of Podolsky parameter. The experimental value for Stefan-Boltzmann constant is $\sigma_0=\left(5,670277968\times 10^{-8}\pm4\times 10^{-13}\right)W/(m^2K^4)$ \cite{Mohr}. Since so far we have not detected any sensible deviation from the Stefan-Boltzmann law, the correction  $\delta\sigma\left(m/T\right)$ must be \textit{at most} equal to the experimental error in the Stefan-Boltzmann constant. Our results showed that such a correction depends on the temperature of the black body radiation. Recalling that CMBR shows a very accurate black body spectrum, we set $T=2,725K$ (the temperature of CMBR) and we find that all values for the Podolsky mass such that $m\gtrsim 4,0\, meV$ are compatible with the experimental data.

\section{Final remarks}\label{conclusions}

In this work we studied the Podolsky theory for electromagnetism at finite temperature. We have reviewed the canonical structure of the theory and have noted that even though  both Maxwell and Podolsky Lagrangians are possible choices for the Abelian gauge group, they give rise different physical results. Using imaginary-time formalism we showed, through Partition Function evaluation, that a gas of free Podolsky photons in thermodynamic equilibrium is \textit{mathematically equivalent} to a gas formed of non-interacting gases of free Maxwell photons and free Proca bosons in thermal equilibrium as well. We argue that this result is only valid for temperatures well bellow the one that corresponds to the electron rest energy. If the temperature is raised enough, we should take into account fermion pair creation \cite{Partovi}. Of course, this can only be accomplished in Podolsky QED. We also showed that Podolsky electromagnetism induces a  modification in Stefan-Boltzmann Law. Accordingly to Podolsky theory at finite temperature, the existing corrections to the Stefan-Boltzmann law vanish as the Podolsky mass goes to infinity recovering, in this approximation, the original law as expected.  Using experimental data for Stefan-Boltzmann constant and the temperature of the CMBR we set a thermodynamical limit to the Podolsky parameter. The reason why we have chosen CMBR temperature is twofold. First of all, CMBR has been called the most accurate black body radiation. Second, it is a temperature well below that associated with the electron mass. Therefore, our results are meaningful in that regime. Our analysis have shown that the Podolsky mass cannot be smaller than approximately $4,0\,meV$, otherwise it would already have been detected in black body radiation experiments. We end this section stating that Podolsky theory remains as a possible choice for the electromagnetic field and only further work in both theoretical and experimental research fields will be able either to  confirm it or to rule it out.

\section{Acknowledgements}\label{acknowledgements}

C. A. B.  and R. B.  thank CNPq for total support. B. M. P. and G. E. R. Z. thank CNPq for partial support.

\end{document}